\documentclass{llncs}
\usepackage{amssymb}
\usepackage{amsmath}
\usepackage{enumerate}
\usepackage{graphicx}
\usepackage[english]{babel}
\usepackage{multirow}
\usepackage{booktabs}
\usepackage{array}
\newcolumntype{L}[1]{>{\raggedright\let\newline\\\arraybackslash\hspace{0pt}}m{#1}}
\newcolumntype{C}[1]{>{\centering\let\newline\\\arraybackslash\hspace{0pt}}m{#1}}
\newcolumntype{R}[1]{>{\raggedleft\let\newline\\\arraybackslash\hspace{0pt}}m{#1}}

\begin{document}

\title{Interpolation of CT Projections by Exploiting Their Self-Similarity and Smoothness}

\author{Davood Karimi  \and Rabab K. Ward}

\institute{Department of Electrical and Computer Engineering \\ University of British Columbia}

\maketitle

\begin{abstract}

As the medical usage of computed tomography (CT) continues to grow, the radiation dose should remain at a low level to reduce the health risks. Therefore, there is an increasing need for algorithms that can reconstruct high-quality images from low-dose scans. In this regard, most of the recent studies have focused on iterative reconstruction algorithms, and little attention has been paid to restoration of the projection measurements, i.e., the sinogram. In this paper, we propose a novel sinogram interpolation algorithm. The proposed algorithm exploits the self-similarity and smoothness of the sinogram. Sinogram self-similarity is modeled in terms of the similarity of small blocks extracted from stacked projections. The smoothness is modeled via second-order total variation. Experiments with simulated and real CT data show that sinogram interpolation with the proposed algorithm leads to a substantial improvement in the quality of the reconstructed image, especially on low-dose scans. The proposed method can result in a significant reduction in the number of projection measurements. This will reduce the radiation dose and also the amount of data that need to be stored or transmitted, if the reconstruction is to be performed in a remote site.

\end{abstract}

\section{Introduction}

There is a great interest in reducing the radiation dose in computed tomography (CT). In practice, this can be done by either lowering the x-ray photon current or reducing the number of projection measurements taken. However, reconstructing a diagnostically useful image from such noisy and/or undersampled measurements is very challenging. As a result, recent years have witnessed a substantial growth in research on more sophisticated image reconstruction and processing algorithms for CT. Most of the published research have focused on iterative reconstruction algorithms \cite{sidky2008} \cite{karimi2016hybrid} \cite{karimi2017sparseview} or on image-domain denoising and restoration algorithms \cite{kang2017deep} \cite{karimi2016coupled} \cite{geyer2015state} \cite{zhang2018sparse} \cite{wu2017iterative} \cite{karimi2016structured}. Comparatively, many fewer studies have been reported on denoising \cite{karimi2016bregman} \cite{karimi2016joint} \cite{karimi2016sinoprojection}, or otherwise improving \cite{karimi2015angular}, the CT projection measurements (i.e., the sinogram). In this paper, we propose a sinogram interpolation algorithm for cone-beam CT. Our algorithm exploits both the smoothness and the self-similarity of the sinogram. We apply the proposed algorithm on simulated and real CT scans and compare it with an algorithm that is based on learned dictionaries.

\section{Materials and Methods}

\begin{figure}[ht]
    \centering
    \includegraphics[width=\textwidth]{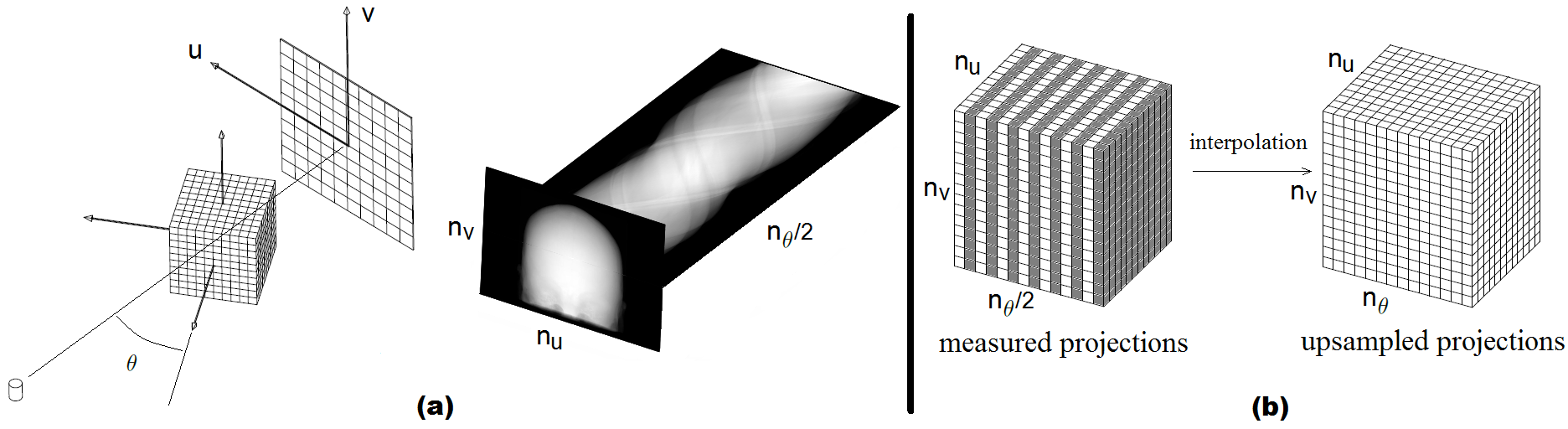}
    \caption{(a) A schematic of cone-beam CT and sinogram of a brain phantom; (b) A schematic representation of the sinogram interpolation problem.}
    \label{fig:cbct_schematic}
\end{figure}

A schematic of the cone-beam CT and a sample sinogram of a brain phantom are shown in Figure \ref{fig:cbct_schematic}(a). The variation of the photon flux incident on the detectors follows a Poisson distribution. However, after the logarithm transformation, the noise follows approximately a Gaussian distribution with signal-dependent variance. Specifically, let us denote the true line integral of the attenuation coefficient along the line from the x-ray source to the detector indexed with $(i,j)$ in the $k^{\text{th}}$ projection with $y_t(i,j,k)$. Then the noisy measurement $y_n(i,j,k) \sim \mathcal{N}(y_t(i,j,k),\sigma_{ijk}^2)$, where $\sigma_{ijk}^2 \propto\exp (y_t(i,j,k))$ \cite{wang2008}.

We assume that only a portion of the desired projections have been directly measured and the rest are to be estimated (i.e., interpolated). Due to the lack of space and in order to simplify the presentation of the algorithm, we focus on the case depicted in Figure \ref{fig:cbct_schematic}(b). Specifically, we assume that half of the $n_{\theta}$ desired projection views have been measured and the remaining half are to be estimated. In other words, the measured sinogram is $y_n \in \mathbb{R}^{n_u \times n_v \times n_{\theta}/2}$ and we would like to estimate the interpolated sinogram $y \in \mathbb{R}^{n_u \times n_v \times n_{\theta}}$. As we will explain later, even though our main goal is sinogram interpolation, the proposed algorithm also has an excellent denoising effect. Therefore, we estimate the full set of $n_{\theta}$ projection views, not only the missing $n_{\theta}/2$ views.

We propose to estimate $y$ by minimizing the following cost function:

\begin{equation} \label{eq:proposed_cost_1}
J(y)= \|My-y_n \|_W^2 +  \lambda_s R_s(y) + \lambda_h R_h(y)
\end{equation}
The first term in $J$ is obtained simply by maximizing the log-likelihood of the measurements. In this term, $M$ is a binary mask matrix that removes from $y$ those projection views that have not been measured and $W$ is a diagonal weight matrix whose diagonal elements are inversely proportional to the measurement variance. The regularization functions $R_s$ and $R_h$ will be explained below, and $\lambda_s$ and $\lambda_h$ are regularization parameters.

\subsection{Regularization in terms of sinogram self-similarity, $R_s$}
\label{sec:R_s}

\begin{figure}[ht]
    \centering
    \includegraphics[width=0.95\textwidth]{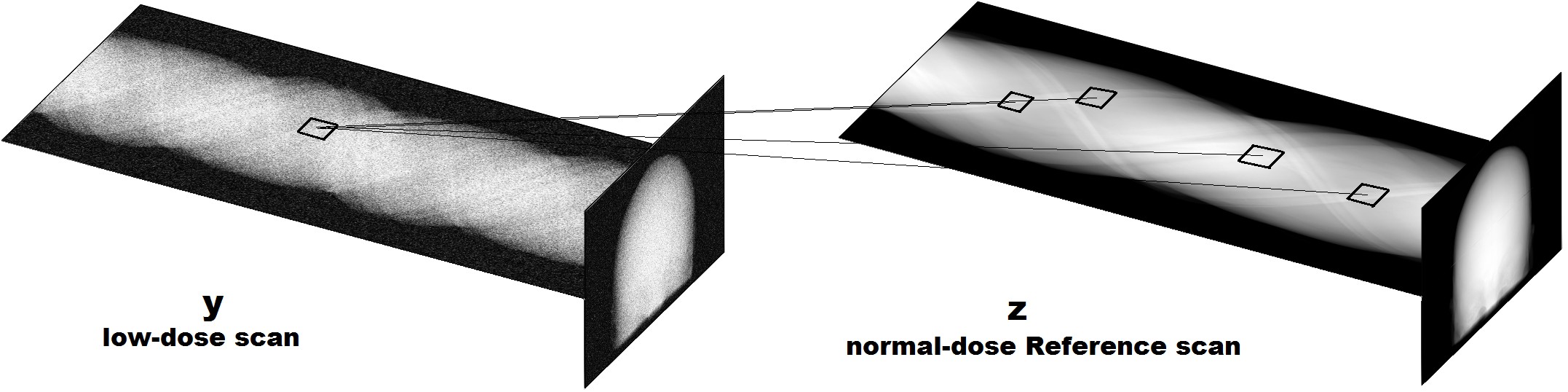}
    \caption{Block matching between the noisy scan to be restored $y$ and the high-dose reference scan $z$.}
    \label{fig:algorithm_schematic}
\end{figure}

In the past decade, patch-based models have emerged as one of the most powerful models in image processing. Much of the research on these models was inspired by the success of the non-local means (NLM) denoising algorithm \cite{buades2005b}. Let us denote the noisy image with $x= x_0+w$, where $x_0$ and $w$ are the true image and the noise, respectively. We also denote the $i^{\text{th}}$ pixel of $x$ with $x(i)$ and a patch centered on $x(i)$ with $x[i]$. NLM estimates the value of $x_0(i)$ as follows:

\begin{equation} \label{eq:nlm}
\widehat{x_0}(i)= \sum_{j=1}^N \frac{G_a(x[j]-x[i])}{\sum_{j=1}^N G_a(x[j]-x[i])} x(j)
\end{equation}

\noindent where $G_a$ denotes a Gaussian kernel with bandwidth $a$ and $N$ is the total number of pixels. The intuition behind this algorithm is that patches that are similar must have similar pixels at their centers. This simple idea has led to some of the best image denoising algorithms and has produced state-of-the-art results in many other image processing applications as well. Many studies have suggested solving various types of inverse problems by introducing regularization terms that are motivated by the same idea \cite{peyre2008,zhang2010b}. Because this type of self-similarity is very abundant in sinogram (even more than in natural images, as can be seen in the sample sinogram in Figure \ref{fig:cbct_schematic}(a)), we suggest a similar form for $R_s$:

\begin{equation} \label{eq:r_s} 
\small
R_s(y)= \|y-y^*\|_2^2 \hspace{2mm} \text{where:} \hspace{2mm} y^*(I)= \sum_{(I') \in \Omega_{I}} \frac{G_a(z[I']-y[I])}{\sum_{(I') \in \Omega_{I}} G_a(z[I']-y[I])} z(I')
\end{equation}

\noindent where we have used simplified pixel index notation $I= (i,j,k)$ and $I'= (i',j',k')$. Computation of $y^*$ has a few differences with the basic NLM in \eqref{eq:nlm}. Firstly, we work with 3D blocks instead of 2D patches. We stack the 2D projections to form a 3D image, as shown in Figure \ref{fig:cbct_schematic}(b), and work with small blocks of this image. This will allow us to exploit both the correlation between adjacent pixels within a projection as well as the correlation between pixels in adjacent projection views. Secondly, for computing $y^*(i,j,k)$, we first find a small number of blocks that are very similar to $y[i,j,k]$ and use those blocks only, instead of all blocks in the image. In \eqref{eq:r_s}, $\Omega_{i,j,k}$ denotes the indices of these blocks. Thirdly, unlike \eqref{eq:nlm} where patch similarities in the same image are exploited, in \eqref{eq:r_s} we use a second image, as shown in Figure \ref{fig:algorithm_schematic}. This second image, denoted with $z$ in Equation \eqref{eq:r_s} and Figure \ref{fig:algorithm_schematic} is built by stacking the projections of the normal-dose (i.e., low-noise) scan of a similar object. For instance, this can be a normal-dose prior scan of the same patient or of a different patient. Our experience shows that with a proper choice of $z$, this approach works much better than finding similar blocks from the same noisy scan. Therefore, computation of $R_s$ requires that for each pixel $y(i,j,k)$ we find a set of blocks sufficiently similar to $y[i,j,k]$ in $z$. We use the Generalized PatchMatch algorithm \cite{barnes2010} for this purpose. We should also note that in finding similar blocks and in computing the block differences in \eqref{eq:r_s}, we only include the pixels from projections that have been directly measured.

\subsection{Regularization in terms of sinogram smoothness, $R_h$}

An important characteristic of sinogram is its smoothness. We model this smoothness via the $\ell_1$ norm of the Hessian of the sinogram. This is a generalization of the $\ell_1$ norm of the gradient, also known as total variation (TV). TV is a good model for piecewise-constant images, but for smooth signals such as sinogram, it leads to staircase artifacts and higher-order differentials need to be considered \cite{papafitsoros2014}. Therefore, we suggest the following form for $R_h$:

\begin{equation} \label{eq:junk}
R_h(y)= \sum_{i,j,k} \left( |\nabla_{uv}^2 y(i,j,k)| + |\nabla_{v\theta}^2 y(i,j,k)| + |\nabla_{\theta u}^2 y(i,j,k)| \right)
\end{equation}

In other words, we compute the 2D Hessians in the three orthogonal planes and add their $\ell_1$ norms. We compute $|\nabla_{uv}^2 y(i,j,k)|$ as:

\begin{equation} \label{eq:forwarddifferences}
\begin{aligned}
|\nabla_{uv}^2 y(i,j,k)|&= \left( D_{uu} y(i,j,k)^2+ 2 D_{uv} y(i,j,k)^2 + D_{vv} y(i,j,k)^2\right) ^{1/2} 
\end{aligned}
\end{equation}

\noindent where $D$s are second-order difference operators. Expressions for $|\nabla_{v \theta}^2 y(i,j,k)|$ and  $|\nabla_{\theta u}^2 y(i,j,k)|$ are very similar.

\subsection{Optimization algorithm}

The proposed cost function has this form:

\begin{equation} \label{eq:proposed_cost_2}
J(y)= \sum_{i,j,k} \bigg( \|My-y_n \|_W^2  +  \lambda_s \|y-y^*\|_2^2 + \lambda_h \Big( |\nabla_{uv}^2 y| + |\nabla_{v\theta}^2 y| + |\nabla_{\theta u}^2 y| \Big) \bigg)
\end{equation}

\noindent where we have dropped the pixel indices to simplify the expressions. An estimate of the interpolated (and denoised) projections is obtained as a minimizer of $J(y)$. To perform this minimization, we use the split Bregman iterative algorithm \cite{goldstein2009}. The split Bregman method first converts the unconstrained optimization problem of minimizing $J(y)$ into a constrained problem:

\begin{equation} \label{eq:augmented} \nonumber
\begin{aligned}
&\textbf{minimize   }   \sum \big( \|Mf-y_n \|_W^2 +  \lambda_s \|f-y^*\|_2^2 \big) +  \lambda_h \big( \sum |g_1| + \sum |g_2| + \sum |g_3| \big)  \\
        &\textbf{subject to:  }  f= y,  \,   g_1=\nabla_{uv}^2 y, \, g_2= \nabla_{v\theta}^2 y, \, g_3= \nabla_{\theta u}^2 y \\
\end{aligned}
\end{equation}

This constrained optimization problem is solved via Bregman iteration:

\begin{equation} \label{eq:bregman1} \nonumber
\small
\begin{aligned}
        &\textbf{Initialize:   }   y^0= y_n, f^0= y_n, g_1^0= \nabla_{uv}^2 y_n, g_2^0= \nabla_{v\theta}^2 y_n, g_3^0= \nabla_{\theta u}^2 y_n, \\  
        & \hspace{47pt}  b_1^0= b_2^0= b_3^0= b_4^0= 0 \\
&\textbf{while  } ||y^k-y^{k-1}||^2_2> \epsilon \\
        & \hspace{4pt}  [y^{k+1},f^{k+1},g_1^{k+1}, g_2^{k+1}, g_3^{k+1}]=  \\
        & \hspace{4pt} \operatorname*{arg\,min}_{y,f,g_1,g_2,g_3}  \sum \big( \|Mf-y_n \|_W^2 +  \lambda_s \|f-y^*\|_2^2 \big) +  \lambda_h \sum |g_1| +  \lambda_h \sum |g_2| +  \lambda_h \sum |g_3| \\
       & \hspace{100pt} + \frac{\mu_1}{2} \sum(f-y-b_1^k)^2 + \frac{\mu_2}{2} \sum(g_1- \nabla_{uv}^2 y-b_2^k)^2 \\
       & \hspace{100pt} + \frac{\mu_2}{2} \sum(g_2- \nabla_{v \theta}^2 y-b_3^k)^2 + \frac{\mu_2}{2} \sum(g_3- \nabla_{\theta u}^2 y-b_4^k)^2\\
       & \hspace{4pt}  b_1^{k+1}= b_1^k+ y^{k+1}- f^{k+1} ; \hspace{15pt}  b_2^{k+1}= b_2^k+ \nabla_{uv}^2 y^{k+1}- g_1^{k+1}  \\
       & \hspace{4pt}  b_3^{k+1}= b_3^k+ \nabla_{v \theta}^2 y^{k+1}- g_2^{k+1}  \hspace{15pt}  b_4^{k+1}= b_4^k+ \nabla_{\theta u}^2 y^{k+1}- g_3^{k+1}  \\
&\textbf{end} \\
\end{aligned}
\end{equation}

\noindent In this algorithm, $\mu_i$ are algorithm parameters and $b_i$ are auxiliary variables. The advantage of this reformulation is that the above minimization problem can be split into five smaller problems, one for each of the five variables that can be solved more easily. Minimization with respect to $f$ has a simple closed-form solution. Minimizations with respect to $g_1$, $g_2$, and $g_3$ also have simple soft-thresholding solutions \cite{papafitsoros2014}. Minimization with respect to $y$ is more difficult. We solve this optimization by considering each of the three Hessian terms in turn, each of which can be approximately solved using the Gauss-Seidel method.

The regularization parameters $\lambda_s$ and $\lambda_h$ influence the recovered solution, whereas $\mu_1$ and $\mu_2$ influence the convergence speed. A study of the effect of these parameters is beyond the limits of this paper. A discussion of the role of these parameters for image restoration with second-order TV can be found in \cite{papafitsoros2013,papafitsoros2014}. In our experiments we used $\lambda_h=0.0001$, $\mu_1=\mu_2= 0.001$, and $\lambda_s=1$. % Also note that, as we mentioned above, for the block-matching needed to compute $R_s$ we used the Generalized PatchMatch, which is an iterative algorithm. We perform one iteration of Generalized PatchMatch after every iteration of the main algorithm. In other words, we do not require Generalized PatchMatch to find very good matches in the beginning. Instead, as the main algorithm makes progress, with every iteration of the main algorithm we apply one iteration of Generalized PatchMatch, gradually improving the similarity of the matches found by this algorithm.

\section{Results and Discussion}

As we mentioned above, unlike the iterative reconstruction methods that aim at reconstructing a high-quality image from undersampled projections, our goal is to estimate the ``missing" projections. Therefore, in all of the experiments reported here we used the FDK algorithm for image reconstruction \cite{feldkamp1984} \cite{karimi2015computational}. We compare our algorithm with the dictionary-based sinogram interpolation method proposed in \cite{li2014}, which was shown to be better than spline interpolation.

\subsection{Experiment with simulated data}
\label{sec:simulation}

We first applied our algorithm on scans simulated from a brain phantom, which we obtained from the BrainWeb database \cite{cocosco1997}. We simulated $n_{\theta}$ projections from this phantom, for two values of $n_{\theta}=1440$ and $960$. For each $n_{\theta}$, we first reconstructed the image of the phantom from the full set of $n_{\theta}$ projections and from $n_{\theta}/2$ projections; we denote these images with $x_{n_{\theta}}$ and  $x_{n_{\theta}/2}$, respectively. We then applied the proposed algorithm and the dictionary-based interpolation algorithm to interpolate the subset of $n_{\theta}/2$ projections to generate $n_{\theta}$ projections and reconstructed the image of the phantom from the interpolated projections. We will denote these images with $x^{\text{proposed}}_{n_{\theta}/2}$ and $x^{\text{dict.}}_{n_{\theta}/2}$. We simulated two levels of noise in the projections with different number of incident photons: $N_0=10^6$ and $N_0= 5 \times 10^4$. We will refer to these simulations as low-noise and high-noise, respectively. For both simulations, we assumed the detector electronic noise to be additive Gaussian with a standard deviation of $40$. As the reference scan that we need for block matching for computation of $R_s$, we used the simulated scan of a different brain phantom from the same database.

\begin{figure}
  \centering
  \begin{tabular}{c c c c c}
    \includegraphics[width=23mm]{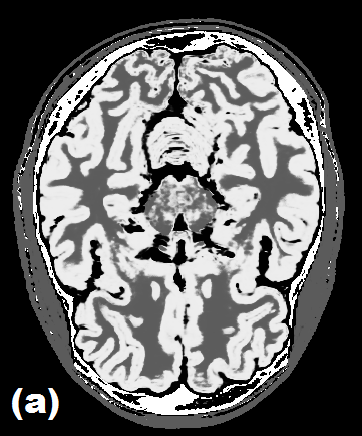} &   
    \includegraphics[width=23mm]{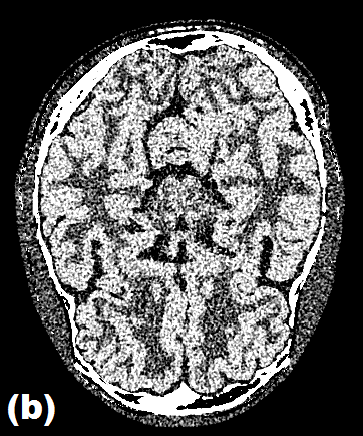} &
    \includegraphics[width=23mm]{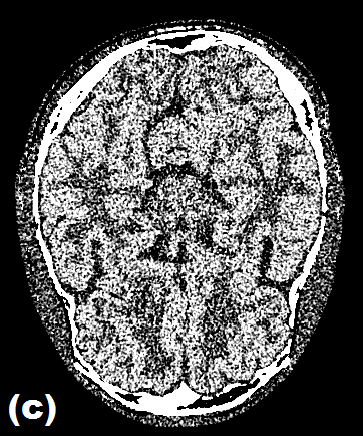} &  
    \includegraphics[width=23mm]{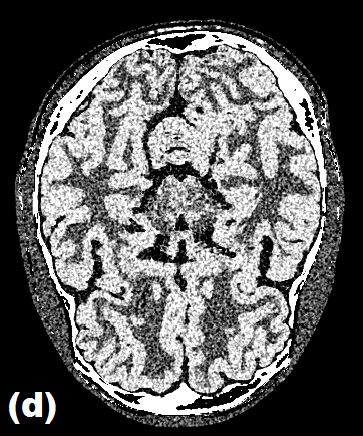} &   
    \includegraphics[width=23mm]{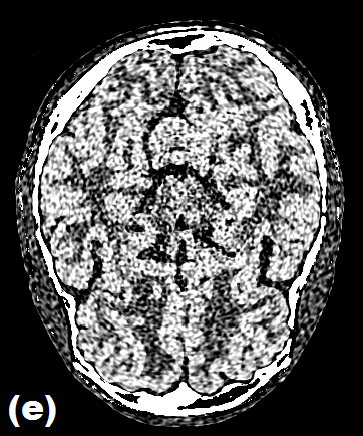} \\ 
  \end{tabular}
  \label{brainslice}\caption{A slice in images of brain phantom reconstructed from the high-noise scan with $n_{\theta}=1440$. (a) the true phantom, (b) $x_{n_{\theta}}$, (c) $x_{n_{\theta}/2}$, (d) $x^{\text{proposed}}_{n_{\theta}/2}$, (e) $x^{\text{dict.}}_{n_{\theta}/2}$.}
  \label{fig:brainslice}
\end{figure}

We compared the reconstructed images with the true phantom image by computing the root-mean-square of error (RMSE) and the Structural Similarity (SSIM) index. The results of this comparison are presented in Table \ref{table:brain_table}. Sinogram interpolation with the propsoed algorithm has resulted in a large improvement in the objective quality of the reconstructed image. The improvement is more substantial in the case of high-noise projections. This is because, as we mentioned above, both regularization terms $R_s$ and $R_h$ have excellent denoising effects. The proposed algorithm has also outperformed the interpolation algorithm based on learned dictionaries. The objective quality of $x^{\text{proposed}}_{n_{\theta}/2}$ is very close to $x_{n_{\theta}}$ in the low-noise case and better than $x_{n_{\theta}}$ in the high-noise case. As shown in Figure \ref{fig:brainslice}, sinogram interpolation with the proposed algorithm has resulted in a substantial improvement in the visual quality of the reconstructed image. Not only artifacts have been significantly reduced, noise has also been decreased substantially.

\begin{table}
  \begin{center}
    \begin{tabular}{| l  l | C{0.9cm} C{0.9cm} C{1.3cm} C{1.2cm} | C{0.9cm} C{0.9cm} C{1.3cm} C{1.2cm} |}
\hline
& & \multicolumn{4}{c}{$n_{\theta}=1440$}  & \multicolumn{4}{c|}{$n_{\theta}=960$} \\
& & $x_{n_{\theta}}$ & $x_{n_{\theta}/2}$ & $x^{\text{proposed}}_{n_{\theta}/2}$ & $x^{\text{dict.}}_{n_{\theta}/2}$ & $x_{n_{\theta}}$ & $x_{n_{\theta}/2}$ & $x^{\text{proposed}}_{n_{\theta}/2}$ & $x^{\text{dict.}}_{n_{\theta}/2}$ \\ \hline
\multirow{2}{*}{Low-noise} & RMSE & 0.104 & 0.140 & 0.111 & 0.128 & 0.127 & 0.154 & 0.130 & 0.138 \\ 
 & SSIM & 0.745 & 0.683 & 0.726 & 0.705 & 0.708 & 0.640 & 0.691 & 0.671 \\ \hline
 \multirow{2}{*}{High-noise} & RMSE & 0.124 & 0.157 & 0.121 & 0.136 & 0.143 & 0.164 & 0.136 & 0.144  \\ 
 & SSIM & 0.710 & 0.642 & 0.715 & 0.686 & 0.688 & 0.625 & 0.690 & 0.662 \\ \hline
    \end{tabular}
  \end{center}
  \caption{Objective quality of the reconstructed images of the brain phantom.}
  \label{table:brain_table}
\end{table}

\vspace{-7mm}

\subsection{Experiment with real CT data}
\label{sec:real_data}

We used two micro-CT scans of a rat for this experiment: (1) a normal-dose scan obtained at normal dose used for routine imaging, (2) a low-dose scan obtained at much reduced dose (by reducing the mAs setting to half of that in routine imaging and using additional copper filtration). Each scan consisted of 720 projections. We reconstructed a high-quality ``reference" image of the rat from all 720 projections of the normal-dose scan. Then, for each of the two scans we selected subsets of $n_{\theta}$ projections for two values of $n_{\theta}=360$ and $180$ and, for each $n_{\theta}$, performed the same analysis as described above for simulation experiment. For the block-matching in the computation of $R_s$ we used the scan of a different rat.

\begin{figure}[ht]
  \centering
  \begin{tabular}{c c c c c}
    \includegraphics[width=23mm]{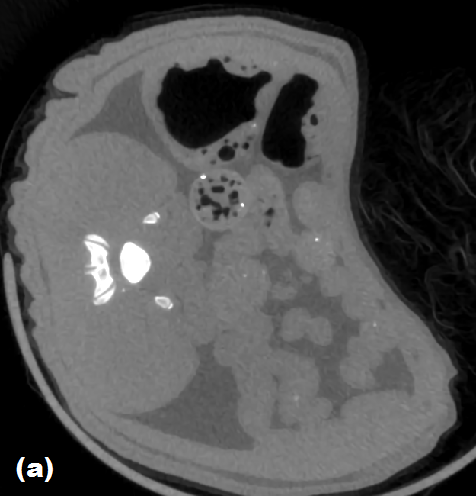} &   
    \includegraphics[width=23mm]{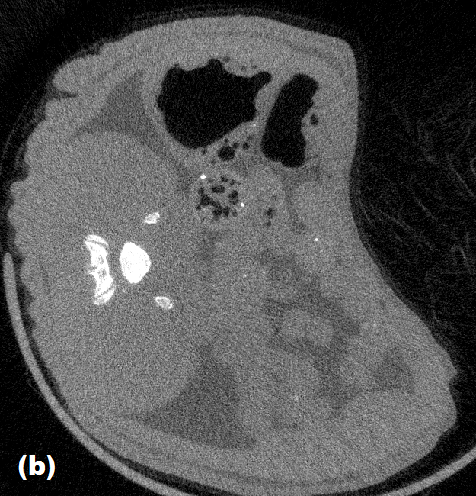} &
    \includegraphics[width=23mm]{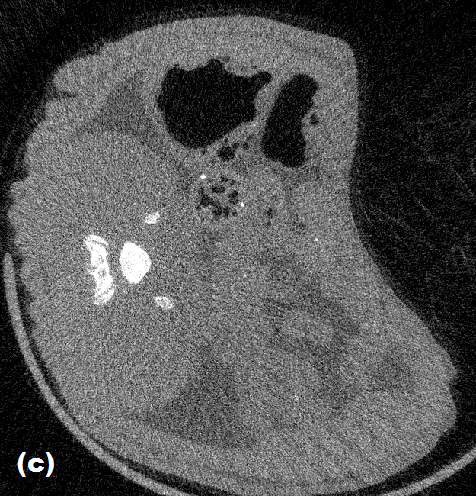} &   
    \includegraphics[width=23mm]{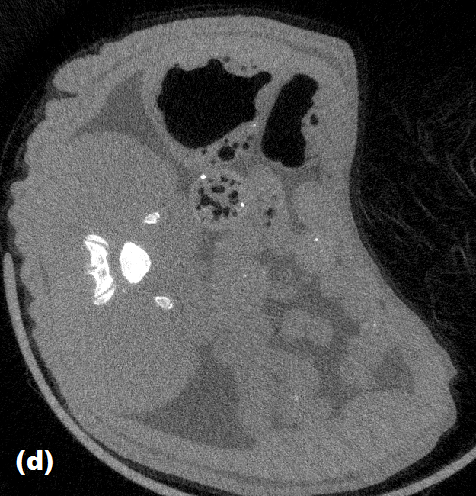} &   
    \includegraphics[width=23mm]{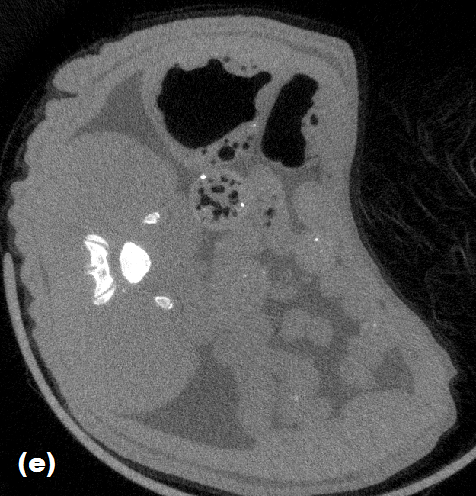} \\ 
  \end{tabular}
  \label{ratslice}\caption{A slice in the images reconstructed from the low-dose scan of the rat with $n_{\theta}=240$. (a) the reference image, (b) $x_{n_{\theta}}$, (c) $x_{n_{\theta}/2}$, (d) $x^{\text{proposed}}_{n_{\theta}/2}$, (e) $x^{\text{dict.}}_{n_{\theta}/2}$.}
  \label{fig:ratslice}
\end{figure}

To assess the quality of the reconstructed images, we computed the RMSE and SSIM with respect to the reference image as well as the contrast-to-noise ratio (CNR). The results of this evaluation have been summarized in Table \ref{table:rat_table}. There is a substantial improvement in image quality metrics as a result of sinogram interpolation with the proposed algorithm. The proposed algorithm has led to much better image quality than the dictionary-based sinogram interpolation. The gain in image quality is more pronounced in the low-dose case. In fact, the objective quality of $x^{\text{proposed}}_{n_{\theta}/2}$ is even better than $x_{n_{\theta}}$ on the low-dose scan. For a visual comparison, Figure \ref{fig:ratslice} shows a slice from the reconstructed images of the rat from the low-dose scan with $n_{\theta}=240$. There is a remarkable improvement in image quality as a result of sinogram interpolation with the proposed algorithm. The visual quality of $x^{\text{proposed}}_{n_{\theta}/2}$ seems to be better than both $x^{\text{dict.}}_{n_{\theta}/2}$ and $x_{n_{\theta}}$.

\begin{table}[ht]
  \begin{center}
    \begin{tabular}{|l  l | C{0.9cm} C{0.9cm} C{1.3cm} C{1.1cm} | C{0.9cm} C{0.9cm} C{1.3cm} C{1.1cm} |}
\hline
& & \multicolumn{4}{c}{$n_{\theta}=360$}  & \multicolumn{4}{c|}{$n_{\theta}=240$} \\
& & $x_{n_{\theta}}$ & $x_{n_{\theta}/2}$ & $x^{\text{proposed}}_{n_{\theta}/2}$ & $x^{\text{dict.}}_{n_{\theta}/2}$ & $x_{n_{\theta}}$ & $x_{n_{\theta}/2}$ & $x^{\text{proposed}}_{n_{\theta}/2}$ & $x^{\text{dict.}}_{n_{\theta}/2}$ \\ \hline
\multirow{3}{*}{Normal-dose} & RMSE & 0.017 & 0.022 & 0.018 & 0.019 & 0.020 & 0.024 & 0.021 & 0.022 \\ 
 & SSIM & 0.654 & 0.611 & 0.640 & 0.636 & 0.632 & 0.582 & 0.624 & 0.617 \\ & CNR & 13.8 & 11.2 & 13.6 & 12.3 & 13.0 & 10.9 & 13.0 & 12.1 \\\hline
 \multirow{3}{*}{Low-dose} & RMSE & 0.019 & 0.024 & 0.019 & 0.020 & 0.022 & 0.025 & 0.022 & 0.023  \\ 
 & SSIM & 0.630 & 0.597 & 0.637 & 0.623 & 0.612 & 0.556 & 0.620 & 0.586 \\ & CNR & 12.3 & 10.5 & 13.0 & 12.1 & 11.7 & 10.2 & 12.2 & 11.5 \\\hline
    \end{tabular}
  \end{center}
  \caption{Objective quality of the reconstructed images of the rat.}
  \label{table:rat_table}
\end{table}

\vspace{-7mm}

\section{Conclusion}

Our experiments show that the proposed sinogram interpolation algorithm can lead to large improvements in image quality. Due to the additional denoising effect of both regularization functions used by the proposed algorithm, this improvement is more significant when applied on low-dose scans. This means that the proposed algorithm is especially well-suited for sinogram restoration in low-dose CT. Our other experiments, not reported in this paper because of space limitations, show that the proposed algorithm can also be used to effectively interpolate the sinogram in more general cases than the case considered in this paper, for example when the angular spacing of the missing projection views is non-uniform or when some detector measurements are corrupted. The experimental results suggest that our proposed method can reduce the number of projection measurements that are needed to reconstruct an image with diagnostic quality. The reduction in the number of projection measurements also means that less data will need to be transmitted in situations where the image reconstruction is performed in a different site.

%\section*{Acknowledgements} 
%
%********************

\bibliographystyle{splncs03}
\bibliography{davoodreferences}

\end{document}